# Carrier Multiplication via Photocurrent Measurements in Dual-Gated MoTe$_2$

*Jun Suk Kim*[†,‡,¶], *Minh Dao Tran*[†,¶], *Sung-Tae Kim*[†,‡], *Daehan Yoo*[§], *Sang-Hyun Oh*[§], *Ji-Hee Kim*[†,‡,*] *and Young Hee Lee*[†,‡,*]

[†]Center for Integrated Nanostructure Physics, Institute for Basic Science (IBS), Suwon 16419, Republic of Korea

[‡]Department of Energy Science, Sungkyunkwan University, Suwon 16419, Republic of Korea

[§]Department of Electrical and Computer Engineering, University of Minnesota, Minneapolis, Minnesota 55455, United States.

**ABSTRACT:**

Although van der Waals-layered transition metal dichalcogenides from transient-absorption spectroscopy have successfully demonstrated an ideal carrier multiplication (CM) performance with an onset of nearly 2E$_g$, interpretation of the CM effect from the optical approach remains unresolved owing to the complexity of many-body electron-hole pairs. We demonstrate the CM effect through simple photocurrent measurements by fabricating the dual-gate p-n junction of a MoTe$_2$ film on a transparent substrate. Electrons and holes were efficiently extracted by eliminating the Schottky barriers in the metal contact and minimizing multiple reflections. The photocurrent was elevated proportionately to the excitation energy. The boosted quantum efficiency confirms the multiple electron-hole pair generation of > 2E$_g$, consistent with CM results from an optical approach, pushing the solar cell efficiency beyond the Shockley–Queisser limit.

One primary concern in single-junction solar cells for energy harvesting is the thermalization loss in semiconductors [1]. The excited electrons subject to incident light thermally relax to the conduction band edge, wasting excess energy above the bandgap as heat with a maximum conversion efficiency of ~34%, which is known as the Shockley–Queisser (SQ) limit [2]. Two approaches have been proposed to overcome the SQ limit: carrier multiplication (CM) by boosting the number of electron-hole pairs [3] and hot-carrier generation by widening the open-circuit voltage [4]. During the CM process, the excess kinetic energy of photoexcited carriers above twice the bandgap can kick out an additional electron-hole pair through an inverse Auger process to multiply the number of electron-hole pairs assisted by a strong Coulomb interaction, which is the key to promoting the CM process in typically confined nano–materials [5–10].

Thus far, intense efforts have been devoted to determining the CM performance using pump-probe techniques. Quantum dots and carbon nanotubes exhibit a diverse CM onset energy of 2.0–3.0$E_g$ and a CM conversion efficiency of 30%–90% [7, 9, 11–13]. Recently, an ideal CM conversion efficiency of ~99% with an onset energy of nearly 2$E_g$ has been reported in transition metal dichalcogenides (TMD) materials such as $MoTe_2$ and $WSe_2$ thin films [9,10]. However, the interpretation of the CM effect based on such an optical approach remains disputable owing to the complexity of the analysis and the possible many-body phenomena [14]. Therefore, it is of critical importance to verify the CM effect through an enhanced quantum yield of photocurrent as a function of photon energy.

Solar cell is a typical device for evaluating the enhanced quantum yield through photocurrent measurement; however, it requires an optimized device architecture with several parameters including an energy band alignment, carrier transport, and choice of interfacial materials [15, 16]. These are the major obstacles to the development of new potential CM materials. Although a two-terminal photocurrent can be simply measured under a strong source-drain voltage [17, 18], this could be further complicated in the presence of a Schottky barrier at the metal contact, with a trap-induced photoconductive hysteresis and strong electric field-induced impact ionization. Therefore, a simple and efficient methodology to prevent such complexities is needed to evaluate the CM performance. In this report, we carefully designed a dual-gate p-n junction structure by stacking ultra-thin two-dimensional vdW layered materials and measured the excitation energy-dependent photocurrent without applying an external source-drain voltage. We are able to achieve a high quantum yield of over 100% with an onset energy of twice the bandgap in $MoTe_2$ thin films.

To minimize the effect of the contact resistance and external electric field in a conventional photoconductive device [Fig. 1(a)], we adopted a dual-gate p-n junction device [Fig. 1(b) and see the Supplemental Material [19], Sec. 1, for details on the fabrication]. Owing to the nearly symmetric ambipolar behavior of the $MoTe_2$ channel [see the Supplemental Material [19], Fig. S2], the Schottky barriers at the metal contacts are efficiently modulated from both gate biases such that the barrier width becomes narrow to allow tunneling to occur. The built-in potential constructed from dual-gate [see the Supplemental Material [19], Fig. S4] effectively extracts the photogenerated carriers in a channel. Our $MoTe_2$ p-n homojunction device consists of an h-BN/$MoTe_2$/h-BN/Gr sandwich structure on a quartz substrate [Fig. 1(c)]. The dual-gate graphite electrodes are separated by ~2 μm to control the local carrier type of the $MoTe_2$ film. This device structure offers three main advantages: i) passivation of a $MoTe_2$ film sandwiched between the top and bottom h-BN thin films to maintain the channel stability, ii) an efficient carrier extraction by eliminating the Schottky barriers in the metal contact, and iii) a minimization of multiple reflections by a transparent substrate. These are the key factors in verifying the CM performance with a high quantum yield. Top-view optical and cross-sectional TEM images clearly identified the stacked heterostructure (top h-BN/$MoTe_2$/bottom h-BN/graphite) of the device [Fig. 1(d) and (e)].

The photo-generated carriers were extracted at zero external bias in the presence of a dual gate across the junction. The contour plot of the photocurrent manifests distinct regions [Fig. 2(a)], i.e., a high photocurrent with a p-n junction at different polarized gate biases and a low photocurrent at the same polarized gate biases. The photocurrent is at maximum and saturated at above $V_{G1}$ (or $V_{G2}$) = 2.0 eV. This is explained by the modulation of the Schottky barrier height with the gate bias, which is nearly zero at a positive gate bias of near $V_G$ = 2.0 eV [inset of Fig. 2(a) and see the Arrhenius plot in Fig. S5]. Spatially resolved photocurrent mapping was applied to confirm the zero Schottky barrier height and estimate the diffusion length of the photo-generated carriers with a scanned step of 200 nm at $V_{DS}$ = 0 V [Fig. 2(b)]. The photocurrent is maximum at the center of the $MoTe_2$ channel and minimum at the contact regions, in contrast with a previous report in which the highest photocurrent was observed with the Schottky barrier at near the contact regions [20]. This confirms the zero Schottky barrier height in the presence of a dual gating. The photocurrent was nearly negligible at the gate electrode region. The diffusion length of the photogenerated carriers is ~ 1 μm, similar to a previous report on few-layer $WSe_2$ [21].

The photocurrent of the dual-gate device ($V_{G1} = -V_{G2} = 2.0$ V) was measured at different light power densities at a fixed photon energy ($E_{ex} = 1.91$ eV) [Fig. 2(c)]. The dark I-V curve of the MoTe$_2$ homojunction device shows an asymmetric diode behavior. The short-circuit current density at 0 V is elevated as the power density increases. The open-circuit voltage is shifted up to ~100 mV at a high power, which could be ascribed to the band filling of the accumulated photocarriers in the MoTe$_2$ channel [21]. The photocurrent was further measured based on the power density and laser energy [Fig. 2(d)]. The photogating effect was nearly negligible with a repeatable photo-response performance with a high on/off ratio, stable on-current level, and fast response time of < 100 μs, regardless of the excitation energy within 2.9 eV. (See the Supplemental Material [19], Fig. S6 and S7) The photocurrent is linearly proportional to the power density within the range of 1–20 mW/cm$^2$ [Fig. 2(e)], ensuring the exclusion of unnecessary many-body interactions in the MoTe$_2$ channel. The difference in the linear slope corresponds to the external quantum yield (EQE) at a given excitation energy.

To obtain the photon-energy dependent EQE, we measured the photocurrent of the device and the corresponding excitation power as functions of the photon energy [Fig. 3(a) and see Supplemental Material [19], Sec. 3, for the detailed measurement]. Subsequently, the EQE is calculated from the ratio of the photocurrent density I to the excitation light intensity P using the equation, $EQE = \frac{I/e}{P/h\nu}$, where $e$ and $h\nu$ denote the elementary charge and photon energy, respectively. The EQE spectrum [Fig. 3(b)] clearly exhibits two pronounced features at approximately 1.8 eV and 2.5 eV, and a small shoulder. The three deconvoluted peaks at near 1.75, 2.0, and 2.5 eV correspond to the A', B', and C excitonic transitions within the visible range in bulk MoTe$_2$ [23]. Note that the observed EQE feature with a transparent quartz substrate is entirely different from that on a silicon substrate because of multiple reflections, which is strongly influenced by the silicon reflection spectrum. (See the Supplemental Material [19], Fig. S10) The absorption spectrum (A) of a MoTe$_2$ flake (thickness of ~18.1 nm) was determined from the experimentally measured transmission spectrum T and reflection spectrum R with $A = 1 - T - R$ [Fig. 3(c) and see the Supplemental Material [19], Sec. 4, for the reflection and transmission measurements]. The measured transmission is almost constant with the increasing photon energy, whereas the reflection steeply doubles from ~25% to ~50%, probably owing to strong light reflection from the oxidized layers (MoO$_x$ and TeO$_x$) on the surface of the MoTe$_2$ flake [24, 25]. Consequently, the absorption is dramatically reduced at high photon energies.

We next calculate the internal quantum efficiency (IQE) by normalizing the EQE spectrum with the absorption spectrum A, *IQE = EQE/A*. Figure 4(a) shows the IQE as a function of excitation energy (or normalized by the MoTe$_2$ bandgap of 0.9 eV [23]). The IQE is gradually doubled from ~15% to ~30%, corresponding to ~100% to ~200% in the normalized IQE (quantum yield) when the excitation photon energy increases from 1.8 eV (~2E$_g$) to 2.6 eV (~2.8E$_g$). A similar change in IQE was observed with a film thickness of 22 nm. (See the Supplemental Material [19], Fig. S14) The doubly enhanced IQE with a photon energy from 2E$_g$ to 3E$_g$ is the fingerprint of CM in our MoTe$_2$ p-n junction device, which is fully consistent with the CM results of a MoTe$_2$ film investigated using optical approaches [9]. The slow upturn of the IQE near the CM onset energy can be attributed to locally induced traps and defects [26].

Finally, we compare the quantum yield of MoTe$_2$ from the photocurrent measurement with the reported CM efficiency for MoTe$_2$ as well as quantum dots using an optical approach. The onset CM energy from the photocurrent measurement is twice the bandgap, congruent with that of the photo-bleaching signal in the pump-probe optical approach [9]. The CM conversion efficiency resides within the range of 95%–99% between photo-bleaching and photo-induced absorption signals from MoTe$_2$, which is still much higher than those of PbS [27] and PbSe quantum dots [28]. This high photocurrent CM efficiency in the MoTe$_2$ p-n junction device is attributed to the small metal contact resistance and short channel length within the diffusion length, allowing for an efficient extraction of the photoexcited carriers. While several challenges remain to be resolved, such as defect passivation, metal contact, and band alignment for electron-hole extraction, this work represents a step forward toward a CM solar cell with 2D materials beyond the SQ limit.

**ACKNOWLEDGMENT**

This work was supported by the Institute for Basic Science (IBS-R011-D1).

*Correspondence to: (J. -H. K.) kimj@skku.edu, (Y. H. L.) leeyoung@skku.edu.

¶J. S. K. and M. D. T. contributed equally to this work

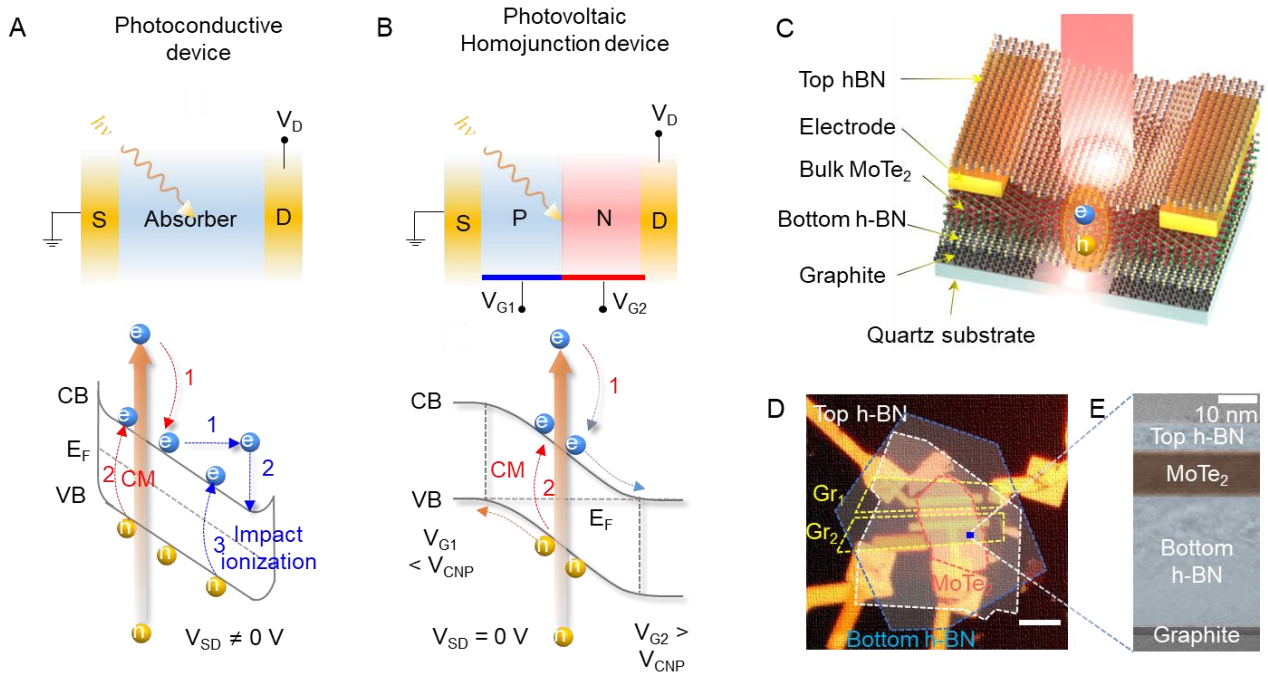

**Fig. 1. Schematic illustration of two photocurrent approaches for exploring carrier multiplication.** (a) CM by twice the bandgap with high source-drain voltage to provoke additional impact ionization process in photoconductive device. Electrons in conduction band gain high kinetic energy by strong electric field (II$_1$), then interact with bound electrons and consequent creation of additional electron-hole pairs (II$_2$). (b) CM with dual-gate to construct p-n junction. Multiple electron-hole pairs generated through CM process are separated and extracted by built-in potential constructed using p-n junction. (c) Schematic of our MoTe$_2$ homo p-n junction device, consisting of top h-BN/MoTe$_2$/bottom h-BN/Gr on quartz substrate. MoTe$_2$ flake with 17.8 nm thickness is encapsulated between top and bottom h-BN flakes. (d) Photograph of MoTe$_2$ device fabricated on a quartz substrate. (e) Cross-sectional TEM image of vertically stacked MoTe$_2$ device.

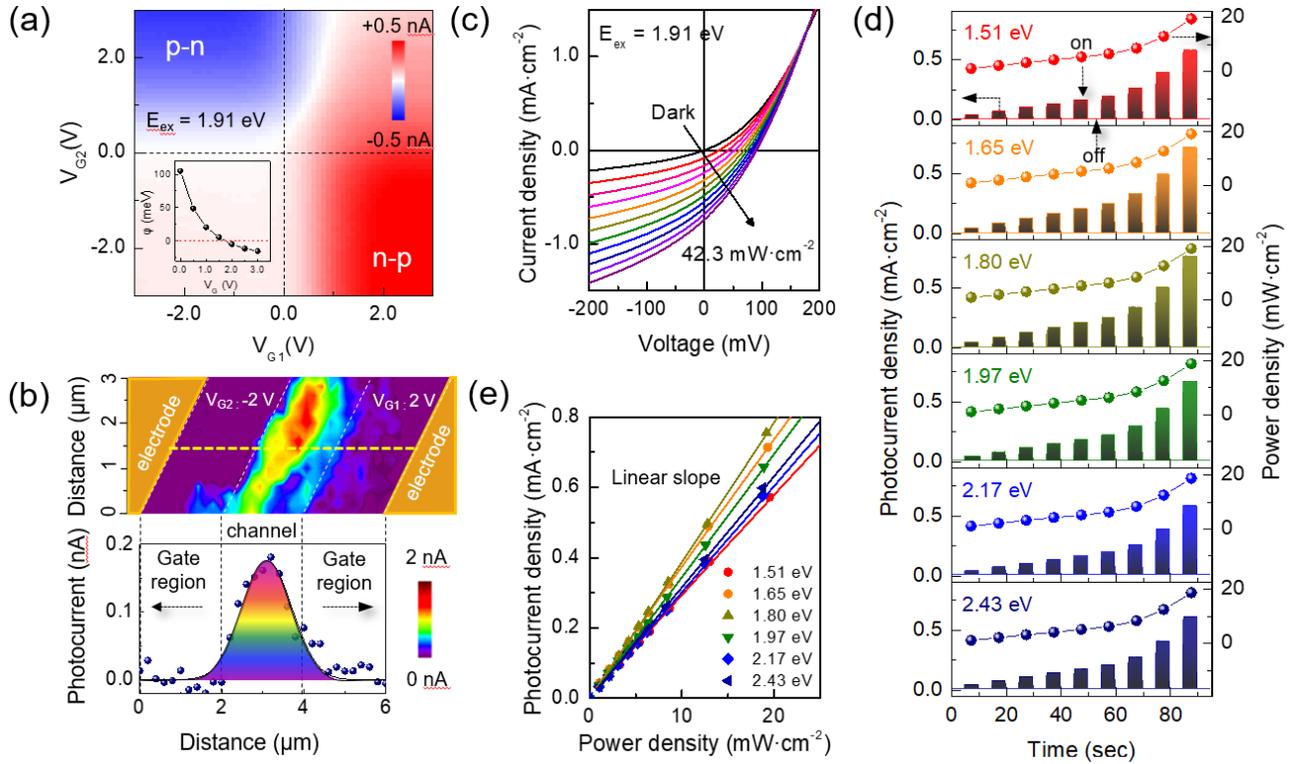

**Fig. 2. Photovoltaic characteristics of MoTe$_2$ p-n junction device.** (a) Contour plot of photocurrent at V$_{DS}$ = 0 V as a function of independent dual gate biases with an excitation energy of 1.91 eV (16 mW/cm$^2$). Inset shows effective Schottky barrier height at various gate biases. (b) (Top) Spatially resolved photocurrent map measured under V$_{G1}$ = -V$_{G2}$ = 2 V with excitation energy of 2.3 eV (25 mW/cm$^2$). (Bottom) Photocurrent profile along dashed yellow line. (c) Power-dependent I-V characteristics under light illumination of 1.91 eV. (d) Power-dependent photo-response with various photon energies. (e) Linear plot of photocurrent versus excitation power obtained from (d). Solid lines indicate linear fit of each energy.

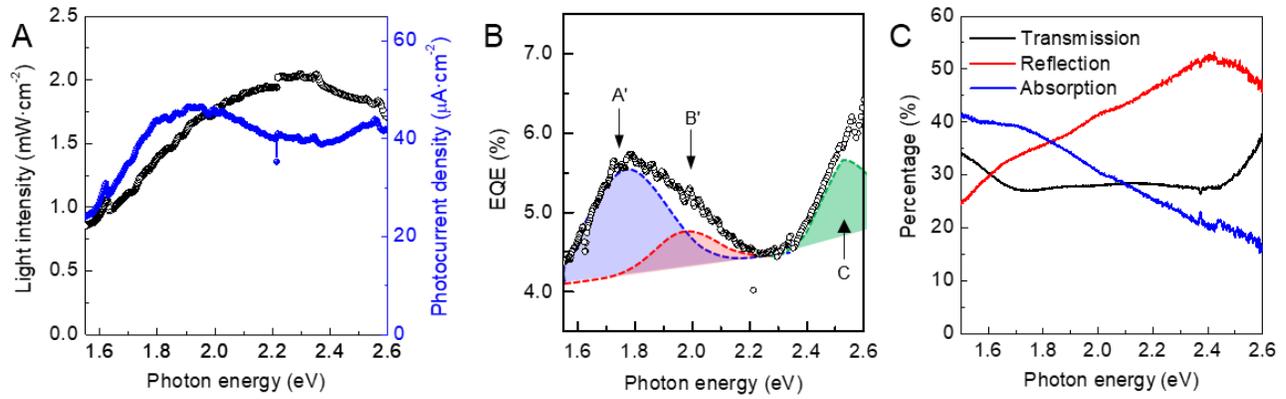

**Fig. 3. Energy-dependent photocurrent of MoTe$_2$ p-n junction device and absorption.** (a) Power density and photocurrent density with photon energy. (b) External quantum efficiency (EQE) of MoTe$_2$ homojunction device. A', B', and C exciton peaks of MoTe$_2$ are shown. (c) Transmission (black), reflection (red), and absorption (blue) spectra of MoTe$_2$ film with a thickness of 18.1 nm.

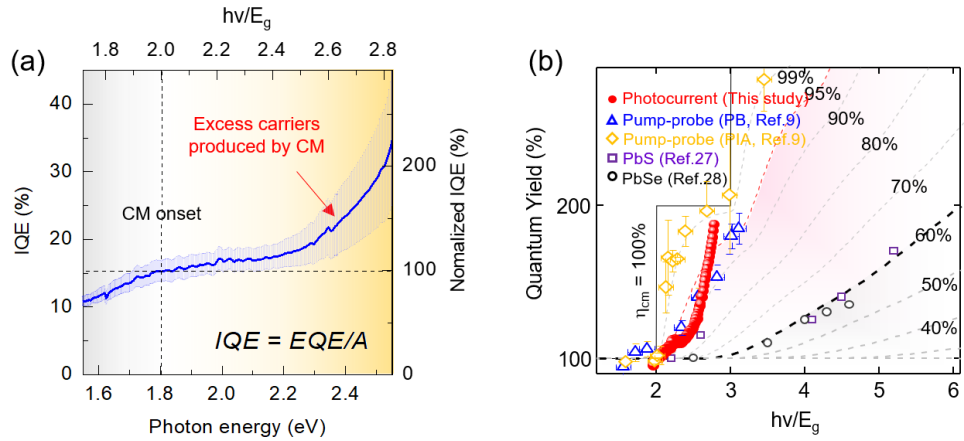

**Fig. 4. IQE of MoTe$_2$ p-n junction device.** (a) IQE with photon energy with error bars representing maximum and minimum values. Error bars were determined by variance in absorption originating from ±2 nm uncertainty of MoTe$_2$ thickness. Dashed line represents CM onset energy of 2E$_g$. (b) Comparison of CM conversion efficiency of MoTe$_2$ measured from photocurrent including previous reports on photo-bleaching (PB) and photo-induced absorption (PIA) in MoTe$_2$ thin film [9] and PbS [27] and PbSe [28] quantum dots measured using optical approach.